\documentclass[12pt]{article}
\usepackage{setspace}
\usepackage[]{authblk}
\usepackage[]{natbib}
\usepackage{graphicx}
\DeclareGraphicsExtensions{.jpg}

\begin{document}


\title{Diffusion-Controlled Quasi-Stationary Mass Transfer for an Isolated Spherical Particle in an Unbounded Medium}
\author{JAMES Q. FENG}
\affil{Boston Scientific Corporation, 3 Scimed Place, Maple Grove, MN 55311, USA\\
james.feng@bsci.com}
\date{}

\maketitle

\begin{abstract}
A consolidated mathematical formulation of the spherically symmetric mass-transfer problem is presented,
with the quasi-stationary approximating equations derived from
a perturbation point of view for the leading-order effect.
For the diffusion-controlled quasi-stationary process, a mathematically complete set of the exact analytical solutions is obtained 
in implicit forms to cover the entire parameter range.  
Furthermore, accurate explicit formulas for the particle radius as a function of time are also constructed semi-empirically
for convenience in engineering practice.
Both dissolution of a particle in a solvent and growth of it by precipitation in a supersaturated environment are considered 
in the present work.
\end{abstract}

\section*{Keywords}
Mass transfer; Diffusion; Dissolution; Precipitation; Particle; Mathematical analysis

\section{Introduction}
Many technological applications involve mass transfer with respect to particles.   
For fundamental understanding in mathematical terms, 
the problem of mass transfer to and from a particle is typically treated as an isolated sphere 
with time-dependent radius in a continuous medium of infinite extent as a consequence of heat-mass transfer
\citep[cf. ][]{scriven59, duda69, cable87, rice06}.
When the attention is focused on the dissolution (or growth by precipitation) of solid particles in liquids, 
as especially important in pharmaceutical dosage form development, 
the mass transfer problem may often be simplified by ignoring the effects of convection and phase-change heating  
such that the governing equations become linear with the ``quasi-stationary'' treatment.
Thus, the mathematical problem is tractable for deriving analytical solutions as usually desired for engineering evaluations.
Moreover, when the mass transfer is mainly limited by the diffusion process rather than the rate of phase change, 
as often to be the case in several realistic applications, the solute concentration at the particle-medium interface can be assumed to
take a constant value of the so-called solubility.
Then the mathematical problem physically describes a diffusion-controlled mass transfer process,
with all the boundary conditions given in the form of Dirichlet type. 
Despite the efforts of many authors over years,
the mathematical analyses of this relatively simplified diffusion-controlled quasi-stationary mass transfer problem
have not been thoroughly satisfactory in terms of completeness and clarity.
Basic understanding of the accuracy and validity of some approximation formulas seems to be lacking in the literature. 

The purpose here is to first present a consolidated mathematical formulation of 
the spherically symmetric mass-transfer problem,
then to derive the quasi-stationary approximating equations mainly based on a perturbation procedure for the leading-order effect,
and to provide a complete set of exact analytical solutions for the entire parameter range.
Because the exact solutions can only be written in implicit forms, 
effort in semi-empirical construction of explicit formulas for the particle radius as a function of time 
is also made for convenience in engineering practice.

\section{Problem Formulation}\label{formulation}
The diffusion-controlled mass transfer to and from a spherical solid particle of (a time dependent) radius $\breve{R}$ in an incompressible continuous fluid medium with a constant density $\rho_m$ and a constant diffusion coefficient $D$
is governed by \citep[][p. 557]{bird60}
\begin{equation}\label{diffusion_eq}
\frac{\partial \breve{C}}{\partial \breve{t}}+v\frac{\partial \breve{C}}{\partial \breve{r}} = \frac{D}{\breve{r}^2}\frac{\partial}{\partial \breve{r}}\left(\breve{r}^2 \frac{\partial \breve{C}}{\partial \breve{r}}\right) 
\quad \breve{R} \le \breve{r} < \infty \quad ,
\end{equation}
where $\breve{C}$ denotes the mass concentration of the solute (namely the dissolved solid from the particle), $\breve{t}$ the time, and $\breve{r}$ the radial distance from the center of the sphere.  In an incompressible fluid with a spherically symmetric flow, the radial velocity
is simply 
\begin{equation}\label{continuity_eq}
v = \left(1-\frac{\rho_p}{\rho_m}\right) \, \left(\frac{\breve{R}}{\breve{r}}\right)^2 \, \frac{d\breve{R}}{d\breve{t}} \, ,
\end{equation}
where $\rho_p$ denotes the (constant) solid particle density,
 to satisfy the equation of continuity and to account for the effect of volume change during the solute phase change 
\citep[e.g., ][]{scriven59}.  At the particle surface, the mass balance based on Fick's first law of binary diffusion accounting for the bulk flow effect with the solvent flux being ignored  \citep[][p. 502]{bird60} leads to
\begin{equation}\label{surface_eq}
\frac{d \breve{R}}{d \breve{t}} = \left[\frac{D}{\rho_p (1 - \breve{C}/\rho_m)} \frac{\partial \breve{C}}{\partial \breve{r}}\right]_{\breve{r}=\breve{R}} \, .
\end{equation}

In a diffusion-controlled process, the typical boundary conditions for $\tilde{C}$ are
\begin{equation}\label{bc_eq}
\breve{C} = \breve{C}_S \mbox{ at the moving boundary } 
\breve{r} = \breve{R} \mbox{ ,  and  } \breve{C} = \breve{C}_0 \mbox{ at } \breve{r} = \infty \, ,
\end{equation}
and initial conditions are
\begin{equation}\label{ic_eq}
\breve{C} = \breve{C}_0 \mbox{ and }  \breve{R} = R_0  \mbox{  at } \breve{t} = 0 \, ,
\end{equation}
where $\breve{C}_S$ denotes the solubility (or `saturated mass concentration') of the solute in the fluid medium\footnote{Here the solubility $\breve{C}_S$ is treated as a constant, 
implying that the particle size effect on solubility as may be observed for submicron particles
(often due to significant surface energy influence), is ignored for theoretical simplicity} 
and $\breve{C}_0$ the initial uniform solute concentration.

If we consider $C(r, t)$ as a dimensionless variable $(\breve{C} - \breve{C}_0)/(\breve{C}_S-\breve{C}_0)$, measure length in units of $R_0$ and time in units of $R_0^2/(\pi D)$, the governing equations (\ref{diffusion_eq})-(\ref{ic_eq}) can be written in a nondimensionalized form
\begin{eqnarray}\label{gov_eq}
\frac{\partial C}{\partial t}+\left(1-\frac{\rho_p}{\rho_m}\right) \frac{R^2}{r^2}\frac{dR}{dt}\frac{\partial C}{\partial r} = \frac{1}{\pi r^2}\frac{\partial}{\partial r}\left(r^2 \frac{\partial C}{\partial r}\right) 
\quad R(t) \le r < \infty \quad , \\
\frac{d R(t)}{d t} = \epsilon \left(\frac{\partial C}{\partial r}\right)_{r=R(t)} \, , \\
C(R(t), t) = 1  \mbox{  and  } C(\infty, t) = 0  \, , \\
C(r, 0) = 0  \mbox{ and }  R(0) = 1   \, ,
\end{eqnarray}
where $t \equiv \pi D \breve{t}/R_0^2$, $r \equiv \breve{r}/R_0$, $R \equiv \breve{R}/R_0$, and $\epsilon \equiv (\breve{C}_S-\breve{C}_0)/[\pi \rho_p(1-\breve{C}_S/\rho_m)]$.  In genera equationsl (\ref{gov_eq})-(9) describe 
a free-boundary (or moving-boundary) nonlinear problem, intractable to exact analytical solutions.  
But if $\epsilon$ can be regarded as a small parameter (e.g., $|\epsilon| << 1$), which is generally valid for dilute solutions with materials of low solubility when 
$\breve{C}_S$, $\breve{C}_0 << \rho_p$ and $\breve{C}_S$ $<< \rho_m$, the solutions to those nonlinear equations may be successively approximated by solutions of linear equations following a perturbation procedure.

\section{Exact Quasi-Stationary Solutions}\label{solutions}
Usually with perturbation approximation, the leading-order solution describes the most significant part of the phenomenon under study.  Hence attention here is restricted only to the leading-order solutions. 

For small $\epsilon$, usually corresponding to the situation of relatively low solubility,
(7) indicates that the time variation of $R(t)$ is slow comparing to that of $C$.  Thus, the convection term in the convection-diffusion equation (6) can be neglected when considering the leading-order effect.  In terms of perturbation solutions for small $\epsilon$, 
$C$ can be written in the typical expansion form as
\begin{equation}\label{perturb_def}
C = C^{<0>}+\epsilon C^{<1>}+\epsilon^2 C^{<2>}+ ... \, ,
\end{equation}
where the zeroth-order solution $C^{<0>}$ is the base solution at $\epsilon = 0$ to the zeroth-order equations
\begin{eqnarray}\label{zeroth-order_eq}
\frac{\partial C^{<0>}}{\partial t} = \frac{1}{\pi r^2}\frac{\partial}{\partial r}\left(r^2 \frac{\partial C^{<0>}}{\partial r}\right) 
\quad R \le r < \infty \quad , \\
\frac{d R(t)}{d t} = 0  \, , \\
C^{<0>}(R, t) = 1   \mbox{  and  } C^{<0>}(\infty, t) = 0  \, , \\
C^{<0>}(r, 0) = 0  \mbox{ and }  R(0) = 1   \, .
\end{eqnarray}

Obviously, at zeroth-order the particle radius $R$ is not changing with time; it becomes a diffusion problem on a fixed domain $R \le r < \infty$.  The solution of $C^{<0>}$ to (11), which can also be written as 
\[
\frac{\partial (r C^{<0>})}{\partial t} = \frac{1}{\pi}\frac{\partial^2(r C^{<0>})}{\partial r^2} \, , 
\]
is given by
\begin{equation}\label{zero-order-C}
C^{<0>} = \frac{R}{r}\left(1-\frac{2}{\sqrt{\pi}} \int_0^{(r-R)\sqrt{\pi/(4 t)}} e^{-\eta^2} d\eta\right)   \, .
\end{equation}
Noteworthy here is that treating $R$ as a time-independent ``variable'' in the zeroth-order diffusion equation is consistent with the so-called ``quasi-stationary'' approximation often used in engineering practice (where the convection transport is neglected and the diffusion equation is solved with the particle surface being considered stationary).  Here we ignore the initial condition $R(0) = 1$ and consider $R$ as a variable yet to be determined.

To evaluate the dissolution process of a particle, we need to consider the first-order effect of $\epsilon$ at least for $R(t)$ in (7) where only the zeroth-order $C^{<0>}$ 
given by (\ref{zero-order-C}) is involved, i.e., 
\begin{equation}\label{first-order-R_eq}
\frac{d R(t)}{d t} = \epsilon \left(\frac{\partial C^{<0>}}{\partial r}\right)_{r=R(t)}  = -\epsilon \left(\frac{1}{R(t)}+\frac{1}{\sqrt{t}}\right)   \, .
\end{equation} 

As pointed out by \citet{krieger67}, \citet{chen89}, and more recently \citet{rice06}\footnote{It seems though the analytical solution 
to (\ref{first-order-R_eq}) obtained by \citet{krieger67} was not noticed by \citet{chen89} and \citet{rice06}. 
However, \citet{krieger67} used their ``highly nonlinear'' implicit formula merely to iteratively determine 
the value of diffusion coefficient,
while \citet{chen89} solved the same equation for an analytical solution with the application in drug particle dissolution in mind.
The recent work of \citet{rice06} again obtained ``an exact analytical solution'' by solving the same mathematical problem
although with a slight change in the form of a parameter to account for the ``bulk flow effect'' as in (3)).},
(\ref{first-order-R_eq}) can be rearranged with some variable substitution to have a form of homogeneous ordinary differential equation as
\begin{equation}\label{hode}
\tau \frac{d u}{d \tau} = -\frac{\epsilon (2 - \epsilon)+(u+\epsilon)^2}{u} \mbox{ or } \tau \frac{d\hat{u}}{d\tau} = -\frac{1+\hat{u}^2}{\hat{u}-\hat{\epsilon}}  \, ,
\end{equation}
where $\tau \equiv \sqrt{t}$, $u \equiv R/\tau$, $\hat{\epsilon} \equiv \sqrt{\epsilon/(2 - \epsilon)}$, and $\hat{u} \equiv (u + \epsilon)/\sqrt{\epsilon (2 - \epsilon)}$ (which are mathematically valid for $0 < \epsilon < 2$).  

Straightforward integration of (\ref{hode}) incorporating the initial condition $R(0) = 1$ yields
\begin{equation}\label{hode_solution}
\tau^2 = t =\frac{\exp[\hat{\epsilon}(2 \tan^{-1}\hat{u}-\pi)]}{ \epsilon ( 2 - \epsilon)(1+\hat{u}^2)}
 \quad (\hat{u} \ge \hat{\epsilon}, 0 < \epsilon < 2) \, .
\end{equation}
This is a solution that can only be expressed in an implicit form of $R(t)$ with $\hat{u} = (R/\sqrt{t}+\epsilon)/\sqrt{\epsilon (2 - \epsilon)}$, though a cleaner formula 
than those presented by previous authors \citep{krieger67, chen89, rice06}.   The formula given by (\ref{hode_solution}) suggests 
an easy way of
generating curves for $R$ as a function of $t$ by first selecting a series of values of $\hat{u}$ ($\ge \hat{\epsilon}$) to calculate $t$
(according to (\ref{hode_solution})), 
and then from the relationship $R = ( \hat{u}\sqrt{2 - \epsilon}-\sqrt{\epsilon})\sqrt{\epsilon t}$ to calculate
the corresponding $R(t)$
from the given $\hat{u}$ and known $t$.
Furthermore, as shown by previous authors \citep{chen89, rice06}, the time to complete dissolution $t_0$ when $R = 0$ happens at 
$\hat{u} = \hat{\epsilon}$ and is thus from (\ref{hode_solution}) given by
\begin{equation}\label{tcd}
t_0 =\frac{\exp[\hat{\epsilon}(2 \tan^{-1}\hat{\epsilon}-\pi)]}{ \epsilon ( 2 - \epsilon)(1+\hat{\epsilon}^2)} 
= \frac{\exp[\hat{\epsilon}(2 \tan^{-1}\hat{\epsilon}-\pi)]}{2 \epsilon}   \, .
\end{equation}

Of mathematical interest, it would be worthwhile to mention that a seemingly different approach to the same mathematical problem was presented by \citet{duda69} who were able to obtain the leading-order quasi-stationary solution of $R(t)$ in an explicit form.   With their approach, the moving interface is immobilized by introducing a boundary-fitted coordinate mapping 
\begin{equation}\label{mapping}
\hat{r} \equiv \frac{r}{R}  \, .
\end{equation}
They also converted the equation system (6)-(9) into one in terms of a compound variable 
\begin{equation}\label{new_variable}
\hat{C} \equiv \hat{r} C  \mbox{  and  } \hat{C} = \hat{C}^{<0>} + \epsilon \hat{C}^{<1>} + ... 
\end{equation}
such that they arrived at
\begin{eqnarray}\label{new_diffusion}
\frac{\partial (\hat{C}^{<0>})}{\partial t} = \frac{1}{\pi}\frac{\partial^2(\hat{C}^{<0>})}{\partial \hat{r}^2} 
\quad 1 \le \hat{r} < \infty \quad , \\
\frac{d R(t)}{d t} = \epsilon \left[\frac{1}{\hat{r} R(t)} \left(\frac{\partial \hat{C}^{<0>}}{\partial \hat{r}} - \frac{\hat{C}}{\hat{r}}\right)\right]_{\hat{r} = 1} = \frac{\epsilon}{R(t)}\left[\left(\frac{\partial \hat{C}^{<0>}}{\partial \hat{r}}\right)_{\hat{r}=1}-1\right]  \, , \\
\hat{C}^{<0>}(1, t) = 1   \mbox{  and  } \hat{C}^{<0>}(\infty, t) = 0  \, , \\
\hat{C}^{<0>}(\hat{r}, 0) = 0  \mbox{ and }  R(0) = 1   \, .
\end{eqnarray}
Similar to (15), the solution of $\hat{C}^{<0>}$ for (22) is then 
\begin{equation}\label{new_solution}
\hat{C}^{<0>} = 1-\frac{2}{\sqrt{\pi}} \int_0^{(\hat{r}-1)\sqrt{\pi/(4 t)}} e^{-\eta^2} d\eta \mbox{  and thus  } \left(\frac{\partial \hat{C}^{<0>}}{\partial \hat{r}}\right)_{\hat{r}=1} = \frac{-1}{\sqrt{t}} \, .
\end{equation}
Therefore, (23) yields
\begin{equation}\label{new_solution_R}
R(t) = \sqrt{1- 2 \epsilon\left(2 \sqrt{t}+t\right)} \, ,
\end{equation}
which is indeed a clean explicit formula.  With this, \citet{duda69} could also carry out derivations of subsequent higher-order perturbation solutions, which would be very difficulty, if not  impossible, with the implicit formula (\ref{hode_solution}).
Now the question is why we can have two apparently different solutions (\ref{hode_solution}) 
and (\ref{new_solution_R}) for the same order of approximation.
To make sure (\ref{hode_solution}) and (\ref{new_solution_R}) are reasonably equivalent mathematical solutions, it is helpful to check the numerical values with each of 
the formulas.
Assuming $\epsilon = 0.1$ (then $\hat{\epsilon}=0.22942$) and $\hat{u}=1$, (\ref{hode_solution}) yields 
$t = 1.83532$ and then $R = 0.45504$ but (\ref{new_solution_R}) gives $R = 0.30173$.  
At $\hat{u}=\hat{\epsilon}=0.22942$, (\ref{tcd}) predicts the time to complete dissolution $t_0=2.69711$ when $R \to 0$, 
but the time to complete dissolution based on (\ref{new_solution_R}) would be $[\sqrt{1+1/(2\epsilon)}-1]^2$ $=2.10102$.
Thus, we see that (\ref{new_solution_R}) cannot be the same as (\ref{hode_solution}) at least for $\epsilon = 0.1$.   

A careful examination of the derivation of (22), however, reveals an error \citep[which was somehow not corrected 
by those authors even in several follow-up publications, e.g.,][]{duda71, vrentas83}.  
The correct expression of (22) should be
\begin{equation}\label{new_diffusion_corrected}
\frac{\partial (\hat{C}^{<0>})}{\partial t} = \frac{1}{\pi R^2}\frac{\partial^2(\hat{C}^{<0>})}{\partial \hat{r}^2} 
\quad 1 \le \hat{r} < \infty \quad .
\end{equation}
Therefore, we should have
\begin{equation}\label{new_solution_corrected}
\hat{C}^{<0>} = 1-\frac{2}{\sqrt{\pi}} \int_0^{(\hat{r}-1) R \sqrt{\pi/(4 t)}} e^{-\eta^2} d\eta \mbox{  and  } \left(\frac{\partial \hat{C}^{<0>}}{\partial \hat{r}}\right)_{\hat{r}=1} = \frac{-R(t)}{\sqrt{t}} \, ,
\end{equation}
which leads to the same equation as (\ref{first-order-R_eq}) and solution as (\ref{hode_solution}) rather than (\ref{new_solution_R}).  
Thus, the same leading-order result can be obtained via seemingly different treatments.
For describing the quasi-stationary dissolution process, (\ref{hode_solution}) should be taken as the correct leading-order solution 
(for $0 < \epsilon < 2$).

It might be noted that so far consideration is only given to the case of $0 < \epsilon < 2$, which describes the 
quasi-stationary dissolution process of a spherical particle.  
Mathematically, solution also exists for the case of $\epsilon < 0$ as well as $\epsilon \ge 2$  in (\ref{first-order-R_eq}).
From a physical point of view, the case of $\epsilon < 0$ in (\ref{first-order-R_eq}) 
describes the inverse process of precipitation growth of a spherical particle,
i.e., when $\breve{C}_0 > \breve{C}_S$ corresponding to the situation of particle growth in a supersaturated solution.  
Somehow, the exact analytical solution to (\ref{first-order-R_eq}) for $\epsilon < 0$ does not seem to have been presented in published literature, 
unlike the case of $\epsilon > 0$.  Here it is derived to complete the mathematical solution for (\ref{first-order-R_eq}).  
With $\epsilon < 0$, (\ref{hode}) must be replaced by
\begin{equation}\label{hode2}
\tau \frac{d u}{d \tau} = \frac{[-\epsilon (2 - \epsilon)]-(u+\epsilon)^2}{u} \mbox{ or } \tau \frac{d\tilde{u}}{d\tau} = \frac{1-\tilde{u}^2}{\tilde{u}+\tilde{\epsilon}}  \, ,
\end{equation}  
where $\tilde{\epsilon} \equiv \sqrt{-\epsilon/(2 - \epsilon)}$, and $\tilde{u} \equiv (u + \epsilon)/\sqrt{-\epsilon (2 - \epsilon)}$.
The solution to (\ref{hode2}) is then
\begin{equation}\label{hode2_solution}
\tau^2 = t =\frac{1}{[- \epsilon (2 - \epsilon)] (\tilde{u}^2-1)}\left(\frac{\tilde{u}+1}{\tilde{u}-1}\right)^{\tilde{\epsilon}} 
\quad (\tilde{u} > 1, \epsilon < 0) \, ,
\end{equation}
which appears to be quite different from (\ref{hode_solution})\footnote{Actually if the identity 
$\ln[(\tilde{u}+1)/(\tilde{u}-1)] = 2 \mbox{ coth}^{-1} \tilde{u}$ is used, we can have (\ref{hode2_solution})
in a similarly looking form to (\ref{hode_solution}) as
\[
\tau^2 = t =\frac{\exp\left(2 \tilde{\epsilon}  \mbox{ coth}^{-1} \tilde{u}\right)}{[- \epsilon (2 - \epsilon)] (\tilde{u}^2-1)}
\quad (\tilde{u} > 1, \epsilon < 0) \, 
\]
where coth$^{-1}$ denotes the inverse hyperbolic cotangent function.}.
As with the dissolution case, curves of $R(t)$ can easily be generated by 
first selecting a series of values of $\tilde{u}$ ($> 1$) to calculate $t$ from (\ref{hode2_solution}), 
and then from the relation $R = ( \tilde{u}\sqrt{2 - \epsilon}+\sqrt{-\epsilon})\sqrt{-\epsilon t}$ to calculate $R(t)$, 
for the particle growth case.

Although the quasi-stationary diffusion-controlled mass transfer problem can be treated as 
a leading-order perturbation problem, it is noteworthy that the perturbation procedure is only performed
on the convection-diffusion equation (6) in terms of the small parameter $\epsilon$ such that 
the neglection of the convection term can be justified in the zeroth-order equation.
In fact, the convection term may naturally disappear in (6) when $\rho_p/\rho_m \to 1$ 
(which might in fact be a quite reasonable assumption especially for a solid solute particle either dissolving in a liquid solvent 
or growing by precipitation in a supersaturated liquid solution, because 
the density of solid solute is usually not much different from that of liquid solution, 
unlike the situation of a liquid droplet in gas or a gas bubble in liquid
with orders of magnitude of density differences).  
In that case, the perturbation treatment in terms of $\epsilon$ becomes unnecessary; 
the ``leading-order'' solution is the exact solution for any value of $\epsilon$.  
Then, (\ref{hode2}) and (\ref{hode2_solution}) with $\tilde{\epsilon}$ being replaced by $-\tilde{\epsilon}$
are also valid for the case of $\epsilon > 2$ 
but with $R = ( \tilde{u}\sqrt{\epsilon-2}-\sqrt{\epsilon})\sqrt{\epsilon t}$
which becomes zero when $\tilde{u} = \tilde{\epsilon}$, where 
$\tilde{\epsilon} \equiv \sqrt{\epsilon/(\epsilon-2)}$ ($> 1$) and $\tilde{u} \equiv (u + \epsilon)/\sqrt{\epsilon (\epsilon-2)}$,
for $\tilde{u} \ge \tilde{\epsilon}$.  

Special attention though should be paid at $\epsilon = 2$ where
\begin{equation}\label{hode3}
\tau \frac{d u}{d \tau} = -\frac{(u+2)^2}{u}  \, ,
\end{equation}
which has the (implicit) solution
\begin{equation}\label{hode3_solution}
\tau = \frac{\exp[-2/(u+2)]}{u+2}  \mbox{ or } t = \frac{\exp[-4/(u+2)]}{(u+2)^2} \quad (0 \le u < \infty) \, ,
\end{equation}
with $R = u \tau$ $= u \sqrt{t}$.  
As expected, (\ref{hode_solution}) indeed approaches (\ref{hode3_solution}) 
at the limit as $\epsilon \to 2$ by applying L'H\^ospital's rule,
and so does (\ref{hode2_solution}) with $\tilde{\epsilon}$ being replaced by $-\tilde{\epsilon}$ for $\epsilon > 2$ 
by using the relationship $e = \lim_{x \to 0}(1 + x)^{1/x}$.

Thus, we have a complete set of solutions to (\ref{first-order-R_eq}) with  
(\ref{hode_solution}) for $0 < \epsilon < 2$, (\ref{hode2_solution}) for $\epsilon < 0$ and $\epsilon > 2$
(with $\tilde{\epsilon}$ being replaced by $-\tilde{\epsilon}$), 
and (\ref{hode3_solution}) for $\epsilon = 2$,
covering all possible $\epsilon \ne 0$ (whenever ignoring the convection term in (6) is justifiable in practice).
Moreover, the case of $\epsilon = 0$ corresponds to a mathematically trivial solution $R = 1$, 
if desired to be included for completeness.

\section{Approximate Explicit Formulas}\label{formulas}
Due to the awkwardness of practical usage of the implicit solution to (\ref{first-order-R_eq}),
authors \citep[e.g., ][]{chen89, rice06} who derived the analytical solution (for $0 < \epsilon < 2$) 
often also suggested explicit ``approximate solution'' with the quasi-steady-state treatment 
(when the time derivative term in the diffusion equation (11) is also neglected).
Simple as it may appear, however, the explicit quasi-steady-state solution may be found not 
to offer a satisfactory approximation to (\ref{hode_solution}) 
for $\epsilon > 10^{-4}$ (depending on the intended application).
To provide much improved approximate explicit formulas for practically accurate evaluation of $R(t)$, effort is made here 
via a semi-empirical approach.

In view of the physical meanings, 
the two terms on the right side of (\ref{first-order-R_eq}) represent two different aspects of the diffusion process: 
$1/R$ comes from the spherically symmetric steady-state solution of 
the Laplace equation whereas $1/\sqrt{t}$ describes transient diffusion from a planar surface.  
Initially when $\sqrt{t} << R$, the concentration gradient front has not propagated far enough from the particle surface 
to bring the curvature effect out yet and thus the effect of transient diffusion from a planar surface dominates.  
With increasing $t$, at some point $\sqrt{t} >> R$ is expected to happen especially for $|\epsilon| << 1$; 
then the steady-state diffusion term becomes dominant and (16) can be reduced to
\begin{equation}\label{qss_ode}
\frac{d R(t)}{d t} = -\frac{\epsilon}{R(t)}   \, ,
\end{equation} 
which is a so-called ``quasi-steady-state'' approximation commonly seen 
in the literature \citep[e.g., ][]{duda71, vrentas83, chen89, rice06}.  
The quasi-steady-state solution to (\ref{qss_ode}) is simply
\begin{equation}\label{qss_ode_solution}
R(t) = \sqrt{1-2 \epsilon \, t}  \mbox{ for } t >> R^2 \, ,
\end{equation}
which yields noticeable difference from (\ref{hode_solution}) even at $\epsilon = 10^{-4}$. 

On the other hand, for $t << R^2$ we have 
\begin{equation}\label{ode3}
\frac{d R(t)}{d t} = -\frac{\epsilon}{\sqrt{t}} \mbox{ and thus } R(t) = 1 - 2 \epsilon \sqrt{t}   \, .
\end{equation} 
Intuitively then, one might believe that a constructed explicit formula like
\begin{equation}\label{approx_solution}
R(t) = \sqrt{1-2 \epsilon \, t} - 2 \epsilon \sqrt{t}  
\end{equation} 
could provide an improved approximation from the quasi-steady-state result (\ref{qss_ode_solution}) to the exact solution  (\ref{hode_solution}).
Clearly, (\ref{approx_solution}) approaches $R = 1 - 2 \epsilon \sqrt{t}$ as $t \to 0$ (where the `higher-order' term associated with
$t = (\sqrt{t})^2$ becomes negligible). 
 The approximate time to complete dissolution corresponding to (\ref{approx_solution}) is
\begin{equation}\label{approx_tcd}
\tilde{t}_0 = \frac{1}{2 \epsilon (2 \epsilon + 1)} \, .
\end{equation}
For comparison at various values of $\epsilon$, the values of time to complete dissolution are tabulated  in table 1
for $t_0$ from the exact (quasi-stationary) solution (\ref{tcd}), the quasi-steady-state result $1/(2 \epsilon)$, 
and the intuitively constructed model result $\tilde{t}_0$ from (\ref{approx_tcd}).
The consistent improvement of (\ref{approx_tcd}) over $1/(2 \epsilon)$ from (\ref{qss_ode_solution})
in comparison with (\ref{tcd}) is obvious, but not as significant as desired.
Both (\ref{qss_ode_solution}) and (\ref{approx_solution}) underestimate the change of $R$ with $t$.  

\begin{table}
\caption{Comparison of predicted values of time to complete dissolution from the quasi-stationary, quasi-steady-state, and intuitively constructed models, with relative error with respect to $t_0$ given in the parentheses.}
\begin{center}
\begin{tabular*}{0.75\textwidth}{@{\extracolsep{\fill}} l c c c }
\hline
\\
$\epsilon$ & {$t_0$ from (\ref{tcd})} & {$1/(2 \epsilon)$ from (\ref{qss_ode_solution})} & { $\tilde{t}_0$ from (\ref{approx_tcd})} \\
\hline
1 & 0.1039 & {0.5 ($381\%$)} & {0.1667 ($60.4\%$)} \\
0.5 & 0.2984  & {1 ($235\%$)}& {0.5 ($67.6\%$)} \\
0.1 & 2.6971 & {5 ($85.4\%$)} & {4.1667 ($54.5\%$)} \\
0.05 & 6.3622 & {10 ($57.2\%$)} & {9.0909 ($42.9\%$)} \\
0.01 & 40.421 & {50 ($23.7\%$)} & {49.020 ($21.3\%$)} \\
0.005 & 85.876 & {100 ($16.4\%$)} & {99.010 ($15.3\%$)} \\
0.001 & 466.54 & {500 ($7.2\%$)} & {499.00 ($7.0\%$)} \\
0.0005 & 952.01 & {1000 ($5.0\%$)} & {999.00 ($4.9\%$)} \\
0.0001 & 4890.6 & {5000 ($2.24\%$)} & {4999.0 ($2.22\%$)} \\
\hline
\end{tabular*}
\end{center}
\end{table}

Interestingly, (\ref{new_solution_R}) obtained by \citet{duda69} is somehow found to offer 
a better approximation than (\ref{approx_solution}) to the exact quasi-stationary solution especially for $\epsilon \le 0.1$,  
although the its derivation has been shown not quite correct in the mathematical sense.   
For example, the time to complete dissolution predicted with $[\sqrt{1+1/(2\epsilon)}-1]^2$ 
based on (\ref{new_solution_R}) would be $0.0505$, $2.1010$, $37.717$, and $457.23$
respectively for $\epsilon = 1$, $0.1$, $0.01$, and $0.001$.
Moreover, (\ref{new_solution_R}) tends to predict faster dissolution whereas (\ref{approx_solution}) slower.  
This is because (\ref{new_solution_R}) over estimated the flux term associated with $1/\sqrt{t}$ in (26) by mistakenly 
replacing $R$ ($< 1$) with $1$.  However, the effect of $1/\sqrt{t}$ usually 
only dominates for a short time when $t$ is small and $R$ is not too far from unity
especially when $\epsilon << 1$.  
Shown in Fig. 1 is a comparison among 
the exact quasi-stationary solution (18), the quasi-steady-state solution (35), 
and the approximate formulas (27) of \citet{duda69} and (37).  
Even at $\epsilon = 0.01$, the deviation of quasi-steady-state solution (35) from the exact solution 
is still quite significant due to the unaccounted initial effect from the flux term $1/\sqrt{t}$ for small $t$.
The overall improvement of (37) from the quasi-steady-state solution (35) is clear, 
especially for small $t$ (or in general $t < 0.5 t_0$) where the curve of (37) consistenly 
remains close to that of the exact quasi-stationary solution. 

\begin{figure}
\centering
\includegraphics[scale=0.64]{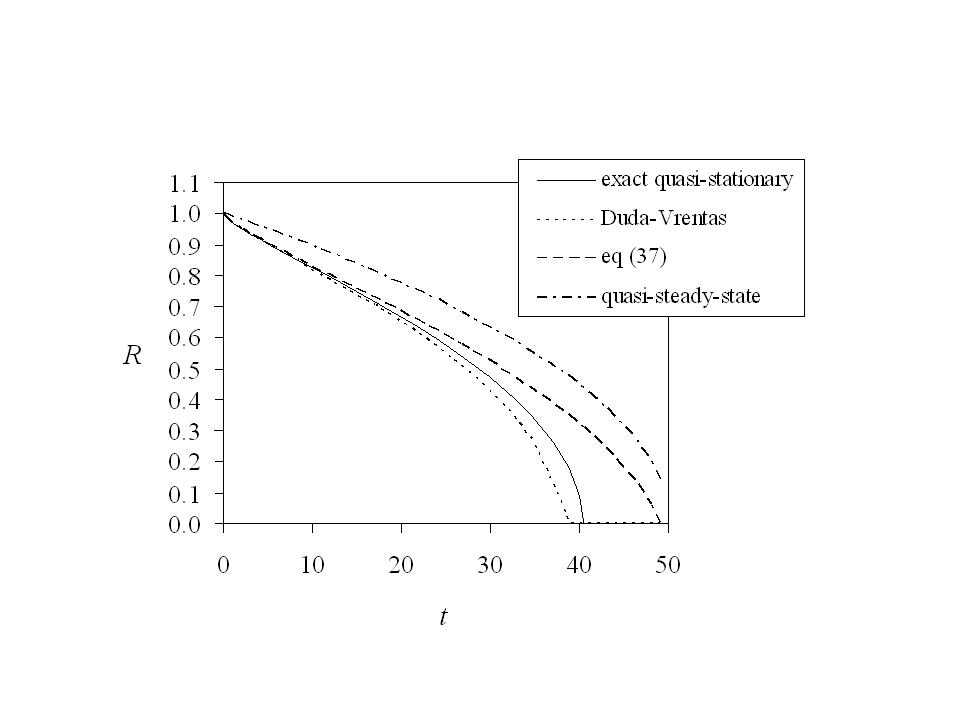}
\caption{Comparison among the exact quasi-stationary solution (18), the quasi-steady-state solution (35), 
and the approximate formulas (27) of \citet{duda69} and (37) for dissolution of particle at $\epsilon = 0.01$.}
\label{fig:fig1}
\end{figure}

In view of Fig. 1, an explicit approximation formula may be constructed semi-empirically by combining 
(\ref{new_solution_R}) and (\ref{approx_solution}) as
\begin{equation}\label{new_approx_solution}
R(t) =  \sqrt{\alpha(\epsilon) \left[1- 2 \epsilon (2 \sqrt{t}+t)\right] + [1-\alpha(\epsilon)] \left(\sqrt{1-2 \epsilon \, t} - 2 \epsilon \sqrt{t}\right)^2}  \, ,
\end{equation}
which with 
\begin{eqnarray}
\alpha(\epsilon) = \left\{
\begin{array}{cc}
  0.781\left[1-1.935/\left(1+1.05 \, \epsilon^{-0.4278}\right)\right]  
\quad (0 < \epsilon < 0.1) \,   \\
  0.0193 (\log_{10}\epsilon)^2-0.2703 \log_{10}\epsilon + 0.095
\quad (0.1 \le \epsilon \le 0.5) \,   
\end{array}
\right . \,
\nonumber
\end{eqnarray}
can consistently produce the value of $R(t)$ very close to that of the exact solution  (\ref{hode_solution}) for any $\epsilon \le 0.5$ 
in the entire range of $0 \le t \le t_0$. 
By the similar token, accurate formulas of $\alpha(\epsilon)$ for $\epsilon > 0.5$ can also be constructed 
but is not attempted here because most practical situations, e.g., in pharmaceutical dosage form development
\citep{curatolo98, kerns08}, typically concern with $\epsilon < 0.1$.

For particle growth in a supersaturated environment, i.e., $\breve{C}_0 > \breve{C}_S$ as in the case of phase separation in
a drying coating, the exact quasi-stationary solution is given by (\ref{hode2_solution}) with $\epsilon < 0$.
Fig. 2 for particle growth shows that comparing to (\ref{hode2_solution}), (\ref{approx_solution}) seems to overestimate the particle growth 
whereas  (\ref{new_solution_R}) of \citet{duda69} and  the quasi-steady-state solution (\ref{qss_ode_solution}) both underestimate it.

\begin{figure}
\centering
\includegraphics[scale=0.64]{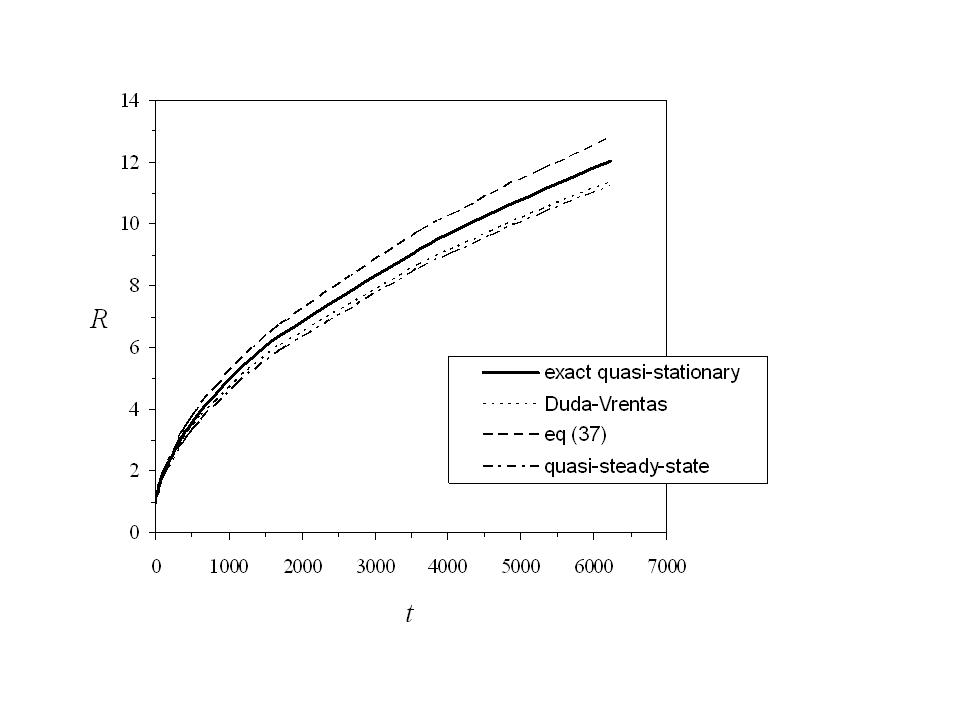}
\caption{Comparison among the exact quasi-stationary solution (31), the quasi-steady-state solution (35), 
and the approximate formulas (27) of \citet{duda69} and (37) for particle growth by precipitation at $\epsilon = 0.01$.}
\label{fig:fig2}
\end{figure}

Based on this observation, a fairly accurate explicit approximate formula to 
the exact quasi-stationary solution (\ref{hode2_solution}) 
may be semi-empirically constructed with the same form as (\ref{new_approx_solution}) but having
\begin{equation}\label{new_alpha}
\alpha(\epsilon) =  0.5381 \left[1-\frac{0.3}{1+|\epsilon|^{-0.6514}}\right]  
\quad (-0.5 \le \epsilon < 0) \,    \, ,
\end{equation}
for evaluating the situation of particle growth.

\section{Concluding Remarks}\label{summary}
Starting from a consolidated mathematical formulation of the spherically symmetric mass-transfer problem,
the quasi-stationary approximating equations can be derived based on
a perturbation procedure for the leading-order effect.
For the diffusion-controlled quasi-stationary process, a mathematically complete set of the exact analytical solutions is obtained 
in implicit forms for consideration of both dissolution of a particle in a solvent and growth of it by precipitation 
in a supersaturated environment.
Understanding the dissolution behavior of solid particles in liquid plays an 
important role in pharmaceutical dosage form development \citep{chen89, rice06}, 
and particle growth by precipitation in a supersaturated environment is relevant to the drug-polymer microsphere formation process
\citep{wu95}
as well as the observed  phase separation process 
during solvent removal in drying of a coating with the drug-polymer mixture \citep{barocas09, richard09}.
The commonly used explicit formula based on the solution with quasi-steady-state approximation is shown 
to provide unsatisfactory accuracy unless the solubility is restricted to very small values (corresponding to $\epsilon < 10^{-4}$).
Therefore, accurate explicit formulas for the particle radius as a function of time 
are also constructed semi-empirically to extend the applicable range at least to $-0.5 \le \epsilon \le 0.5$
for practical convenience.

\section*{Acknowledgment}
The author is indebted to Professor Richard Laugesen of the University of Illinois for his helpful discussions and skillful illustration of mathematical manipulations.
The author also wants to thank Yen-Lane Chen, Scott Fisher, Ismail Guler, Cory Hitzman, Steve Kangas, Travis Schauer,  
and Maggie Zeng of BSC for their consistent support.

\bibliographystyle{}

\end{document}